\documentclass{natureprintstyle}
\usepackage{graphicx}

\pdfoutput=1

\def\aap{Astron. Astrophys.}                
\def\physrep{Phys.~Rep.}   
\def\apj{Astrophys. J.}                 
\def\aj{Astron. J.}                   
\def\apjl{Astrophys. J.}                
\def\araa{Ann. Rev. Astron. Astrophys.}             
\def\jcap{J. Cosmology Astropart. Phys.}
\def\apjs{Astrophys. J.}               
\def\mnras{Mon. Not. R. Astron. Soc.}             
\def\aapr{Astron. Astrophys. Rev.}          
\def\nat{Nature}
\def\beq{\begin{equation}}
\def\eeq{\end{equation}}
\newcommand{\kms}{\,{\rm km\,s^{-1}}}
\newcommand{\msun}{M_{\odot}}

\title{Black hole growth in the early Universe is self-regulated and largely hidden from view}

\author{Ezequiel Treister$^{1,2,3}$, Kevin Schawinski$^{2,4,5}$, Marta Volonteri$^{6}$,
 Priyamvada Natarajan$^{4,5,7,8}$ and Eric Gawiser$^{9}$}

\begin{document}

\maketitle

\begin{affiliations}
\item Institute for Astronomy, 2680 Woodlawn Drive, University of Hawaii, Honolulu, HI 96822
\item Chandra/Einstein Fellow
\item Universidad de Concepci\'{o}n, Departamento de Astronom\'{\i}a, Casilla 160-C, Concepci\'{o}n, Chile
\item Yale Center for Astronomy and Astrophysics, P.O. Box 208121, New Haven, CT 06520.
\item Department of Physics, Yale University, P.O. Box 208121, New Haven, CT 06520.
\item Department of Astronomy, University of Michigan, Ann Arbor, Michigan 48109, USA.
\item Department of Astronomy, Yale University, PO Box 208101, New Haven, CT 06520.
\item Institute for Theory and Computation, Harvard University, 60 Garden Street, Cambrigde, MA 02138.
\item Department of Physics and Astronomy, Rutgers University, 136 Frelinghuysen Road, Piscataway, NJ 08854
\end{affiliations}

\begin{abstract}
The formation of the first massive objects in the infant Universe remains impossible to observe directly and yet 
it sets the stage for the subsequent evolution of galaxies\cite{silk98,barkana01,VHM}. While some
black holes with masses $>$10$^9$M$_{\odot}$ have been detected in luminous quasars less than one billion 
years after the Big Bang\cite{fan01,willott07}, these individual extreme objects have limited 
utility in constraining the channels of formation of the earliest black holes. The initial conditions of black hole seed properties 
are quickly erased during the growth process\cite{volonteri06}. From deep, optimally-stacked, archival  X-ray observations,  we 
measure the amount of black hole growth in $z$=6-8  galaxies (0.7-1 billion years after the Big Bang). Our results imply that 
black holes grow in tandem with their hosts throughout cosmic history, starting from the earliest times.  We find that most copiously 
accreting black holes at these epochs are buried in significant amounts of gas and dust that absorb most radiation except for the highest energy 
X-rays. This suggests that black holes grow significantly more than previously  thought during these early bursts, and 
due to obscuration they do not contribute to the re-ionization of the Universe with their ultraviolet emission. 
\end{abstract}

The {\it Chandra} X-ray observatory is sensitive to photons in the energy range
0.5-8 keV, which in deep extragalactic observations probes predominantly accretion
onto supermassive black holes\cite{brandt05}. Rapidly growing black holes are 
known to be surrounded by an obscuring medium, which can block most of the 
optical, ultra-violet and even soft X-ray photons\cite{lawrence82}. With increasing 
redshift, at the earliest epochs, the photons observed by {\it Chandra} are emitted at 
intrinsically higher energies, and  therefore less affected by such absorption. Current X-ray 
observations have not been able to individually detect most of the first black hole growth events 
at $z$$>$6 (first 950 million years after the Big Bang) thus far, except for the most luminous 
quasars\cite{shemmer06} at $L_X$$>$3$\times$10$^{44}$~erg~s$^{-1}$. While deep X-ray surveys
do not cover enough volume at high redshift, current wide-area studies are simply not deep enough. Hence, the only way to 
obtain a detectable signal from more typical growing black holes is by adding the X-ray emission from a large 
number of sources at these redshifts, which we pursue here.

We start by studying the collective X-ray emission from the most distant galaxies known, 
at $z$$\sim$6\cite{bouwens06}, $z$$\sim$7\cite{bouwens10} and $z$$\sim$8\cite{bouwens10b}, detected by 
the Wide Field Camera aboard the {\it Hubble} Space Telescope. These galaxies are as massive as 
today's galaxies ($~10^9-10^{11} M_\odot$ stellar mass\cite{labbe06}), and they are thus likely to harbor substantial 
nuclear black holes. None of the $z$$>$6  galaxies studied in this work are individually detected in the {\it Chandra} X-ray 
observations. However, we detect significant signals from a stack of 197 galaxies at $z$$\sim$6 in both the soft 
(0.5-2.0 keV; corresponding to 3.5-14 keV in the rest frame) and hard (2-8 keV; rest-frame 14-56 keV) X-ray bands
independently. The detection in the soft band is significant at the 5-$\sigma$ level and implies an average
observed-frame luminosity of 9.2$\times$10$^{41}$~erg~s$^{-1}$, while in the hard band the stacked
6.8-$\sigma$ signal corresponds to an average luminosity of 8.4$\times$10$^{42}$~erg~s$^{-1}$. For the
sample of galaxies at $z$$\sim$7 we obtain 3-$\sigma$ upper limits for the average luminosity in the observed-frame
soft and hard X-ray bands of 4$\times$10$^{42}$~erg~s$^{-1}$ and 2.9$\times$10$^{43}$~erg~s$^{-1}$ 
respectively. Combining the $z$$\sim$7 and $z$$\sim$8 samples the corresponding 3-$\sigma$ upper limits
are 3.1$\times$10$^{42}$~erg~s$^{-1}$ and 2.2$\times$10$^{43}$~erg~s$^{-1}$ in the observed-frame
soft and hard X-ray bands.

A large difference, of a factor of $\sim$9, is found between the stacked fluxes in the soft and hard X-ray bands at
$z \sim 6$. This requires large amounts of obscuring material with high columns ($N_H$$>$1.6$\times$10$^{24}$~cm$^{-2}$) to 
be present in a very high fraction of the accreting black holes in these galaxies, in order to explain the large deficit of soft X-ray 
photons. Since this signal derives from the entire population, these results require that almost all sources are 
significantly obscured. This in turn implies that these growing black holes are obscured along most lines
of sight, as observed in a small subset of nearby objects\cite{ueda07} as well. Such high fraction of obscured sources
at low luminosities is also observed at low redshifts\cite{sazonov07}. This large amount 
of obscuration along all directions absorbs virtually all ultra-violet photons from growing black 
holes. Thus, regardless of the amount of accretion in these sources, these active galaxies cannot 
contribute to the early re-ionization of the Universe. Alternatively, it cannot be claimed that rapid and 
efficient supermassive black hole growth in the high-z Universe is implausible on the basis of any
re-ionization constraints\cite{loeb09}. If most of the high-redshift black
hole growth is indeed obscured as suggested by our work, several current constraints on the lifetime and 
duty cycle of high-z accreting black holes need to be revisited and revised.

Assuming that the X-ray emission is due to accretion onto the central black hole, the accreted black hole mass 
density can be directly derived from the observed X-ray luminosity, as described in the supplementary information.
Extrapolations of Active Galactic Nuclei (AGN) luminosity functions\cite{treister09b} measured at 
significantly lower redshifts, $z$$<$3, are consistent with the observed accreted black hole mass 
density at $z$$>$6, as can be seen in Figure 1. This directly leads to two further conclusions: the 
space density of low luminosity sources,  $L_X$$<$10$^{44}$~erg~s$^{-1}$, does not evolve 
significantly from $z$$\sim$1 to $z$$\sim$6-8, i.e. over more than 5 billion years. Second, at higher 
luminosities, the extrapolation of lower-redshift AGN luminosity functions leads to an 
overestimate of the observed source density in optical surveys\cite{willott10a}. This 
discrepancy can be resolved if the shape of the AGN luminosity function evolves strongly in the sense that there are 
relatively fewer high-luminosity AGN at $z$$>$6 in comparison to the $z$$<$3 
population. Another possibility is that the number of obscured sources, relative to unobscured
quasars, increases with redshift, such that most of the highly obscured systems are systematically 
missed in these optical studies. This is strongly supported by observations of quasars at lower 
redshifts, $z$$<$3\cite{treister10}. We cannot rule out either of these scenarios at present due to the 
relatively small cosmological volume studied, in which the extremely rare high-luminosity AGN are absent.

Our measurements and upper limits for the accreted black hole mass density up to $z$$\sim$8.5 ($\sim$600 million
years after the Big Bang) constrain the nature of black hole growth in the early Universe. Two critical issues
for AGN and the supermassive black holes powering them are how the first black holes formed, and 
how they subsequently grew accreting mass while shining as AGN.  The strong local correlation 
between black hole mass and galaxy bulge mass observed at $z$$\sim$0\cite{ferrarese00,gebhardt00}, is interpreted  
as evidence for self-regulated black hole growth and galaxy-black hole co-evolution\cite{silk98,king03}. This is 
currently the default assumption for most galaxy formation and evolution models\cite{wyithe03,hopkins06}.

The origin of the initial ``seed'' black holes remains an unsolved problem at present. Two channels to form these seeds have 
been proposed: compact remnants of the first stars, the so-called population III stars\cite{madau01}, which generate 
seeds with masses $\sim$10-1,000 M$_\odot$ and from the direct gravitational collapse of gas-rich pre-galactic disks, which leads
 to significantly more massive seeds with masses  in the range $M$$\sim$10$^5$~$M_\odot$\cite{bromm03,lodato06}. By construction, the masses 
of seeds that form from direct collapse are correlated to properties of the dark matter halo and hence properties of the galaxy that will 
assemble subsequently.
 
To interpret our finding, we explore a theoretical framework for the cosmic evolution of supermassive black holes in a $\Lambda$CDM cosmology.
We follow the formation and evolution of black holes  through dedicated Monte Carlo merger tree simulations. Each model is constructed by tracing the merger hierarchy
of  dark matter halos in the mass range $10^{11}-10^{15} {\rm M_\odot}$
backwards to $z=20$, using an extended Press \& Schechter algorithm\cite{VHM}.  The halos are then seeded with black holes and
their evolution is tracked forward to the present time. Following a major
merger (defined as a merger between two halos with mass ratio $>0.1$),
supermassive black holes accrete efficiently an amount of mass that is set by a ``self-regulated" model 
(where the accreted mass scales with the fourth power of
the host halo circular velocity and is normalized to reproduce the
observed local correlation between supermassive black hole mass and velocity
dispersion) or a ``un-regulated" model, where the supermassive black hole simply doubles in mass at each 
accretion episode. See the Supplementary Information for additional details.

Our observational results provide strong support for the existence of a correlation between supermassive black holes and their 
hosts out to the highest redshifts. In Figure 1, we 
compare both unregulated and self-regulated black hole growth models with our observations, and find that physically motivated 
self-regulation growth models are highly favored at all redshifts, even in the very early Universe. Un-regulated models (for instance wherein 
black holes just double in mass at each major merger) are strongly disfavored by the data. This indicates that even in the first episodes of black 
hole growth there is a fundamental link between galaxy and black hole mass assembly. 

As shown in Figure 1, once a standard prescription for self-regulation (as described before) is incorporated, both seed models are consistent 
with our current high-z observations. Detection of an unbiased population of sources at these early epochs is the one metric that we have in the foreseeable future to 
distinguish between these two scenarios for the origin of supermassive black holes in the Universe. In Figure 2, we present the predicted cumulative 
source counts at $z$$>$6 for the models studied here. Based on these models, ultra-deep X-ray and near-infrared surveys covering at 
least $\sim$1 deg$^2$ are required to constrain the formation of the first black hole seeds. This will likely require the use of the next generation 
of space-based observatories such as the James Webb Space Telescope and the International X-ray Observatory.

\begin{addendum}
\item[Supplementary Information] is linked to the online version of the paper at
http://www.nature.com/nature.
 \item[Acknowledgments] We thank Tomo Goto, Meg Urry and Dave Sanders for very insightful
conversations. Support for the work of ET and KS was provided NASA through 
{\it Chandra}/Einstein Post-doctoral Fellowship Awards.  MV acknowledges 
support from the Smithsonian Astrophysical Observatory. PN acknowledges support via 
a Guggenheim Fellowship from the John Simon Guggenheim Foundation.  The work of EG was partially funded
by NSF.
\item[Author Contributions] E.T. started the project, collected the galaxy samples, performed the X-ray
stacking calculations and wrote the majority of the text. K.S. helped to collect the galaxy sample studied 
here, contributed to the conception of the project, the analysis and interpretation of the results. M.V. and P.N. created 
the black hole growth models, computed the contribution of these sources to the re-ionization of the Universe and 
contributed extensively to the theoretical interpretation of the observational results. E.G. developed the optimal 
X-ray stacking formalism and worked with E.T. to implement it on these data. All authors discussed the results and 
contributed to the writing of the manuscript.
\item[Competing Interests] The authors declare that they have no
competing financial interests.
 \item[Correspondence] Correspondence and requests for materials
should be addressed to E. Treister.~(email: treister@ifa.hawaii.edu).
\end{addendum}


\begin{figure}[h!]
\begin{center}
\includegraphics[width=3.5in]{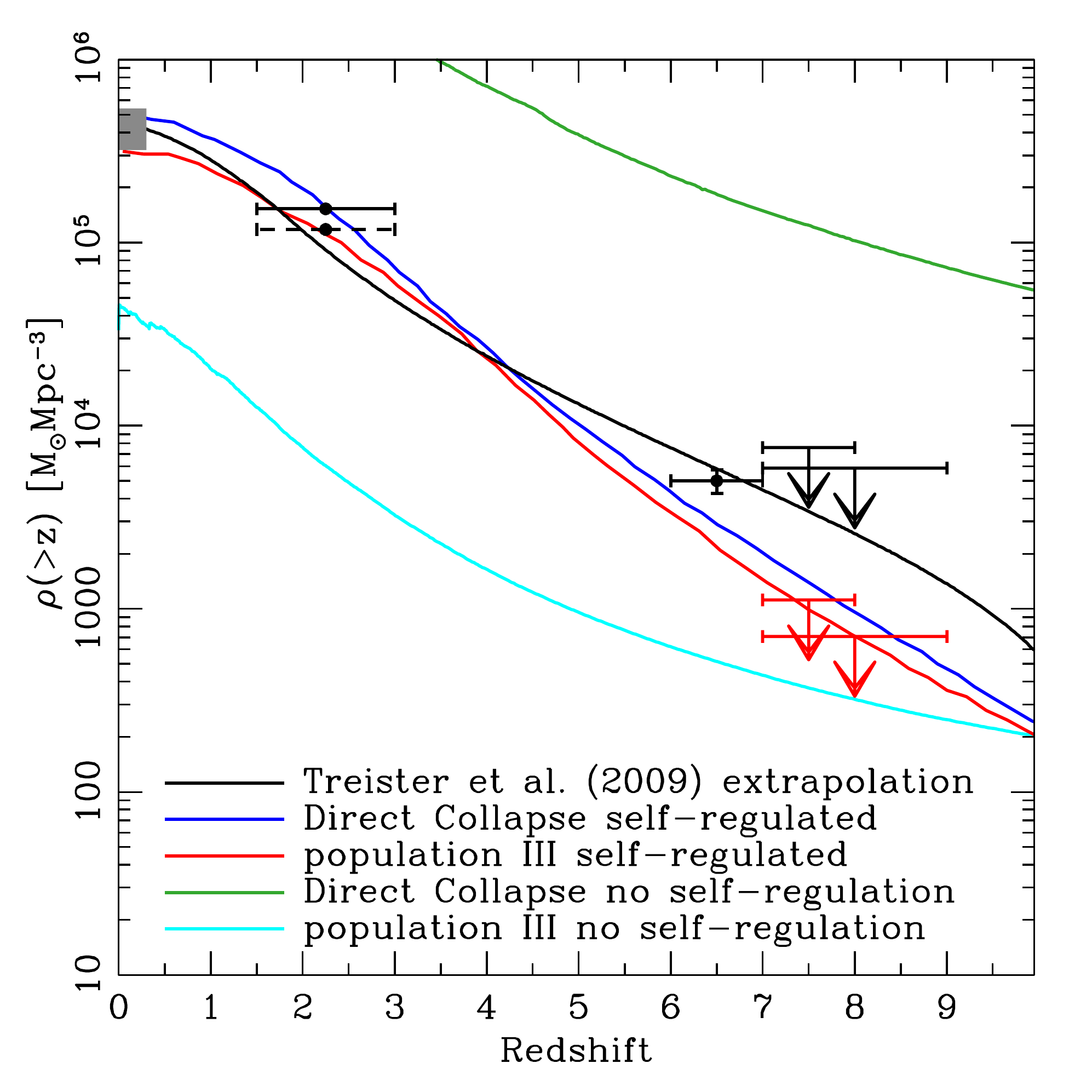}
\end{center}
\vspace{-0.7cm}
\caption{Accreted black hole mass density as a function of redshift. The {\it gray rectangle} shows the  range of values allowed by observations 
of $z$$\simeq$0 galaxies\cite{shankar09}. The data points at $z$$\sim$2 correspond to the values obtained from {\it Chandra} observations of X-ray 
detected AGN and luminous infrared galaxies\cite{treister10b}, while the measurement at $z$$\sim$6 and the upper limits at $z$=7-9 show the 
results described in this work ({\it red} and {\it black} data points from the observed-frame soft and hard X-ray band observations respectively). Vertical error 
bars represent 1 s.d. while the horizontal ones show the bin size. The {\it black} solid line 
shows the evolution of the accreted black hole mass density inferred from the extrapolation of AGN 
luminosity functions measured at lower redshifts\cite{treister09b}. We over-plot the predictions of black hole and galaxy evolution 
models\cite{volonteri10a} for non-regulated growth of Population-III star remnants ({\it cyan line}) and direct-collapse seeds ({\it green}). The {\it red}  
and {\it blue} lines show the predicted BH mass density if self-regulation is incorporated.}
\end{figure}

\begin{figure}[t!]
\begin{center}
\includegraphics[width=3.5in]{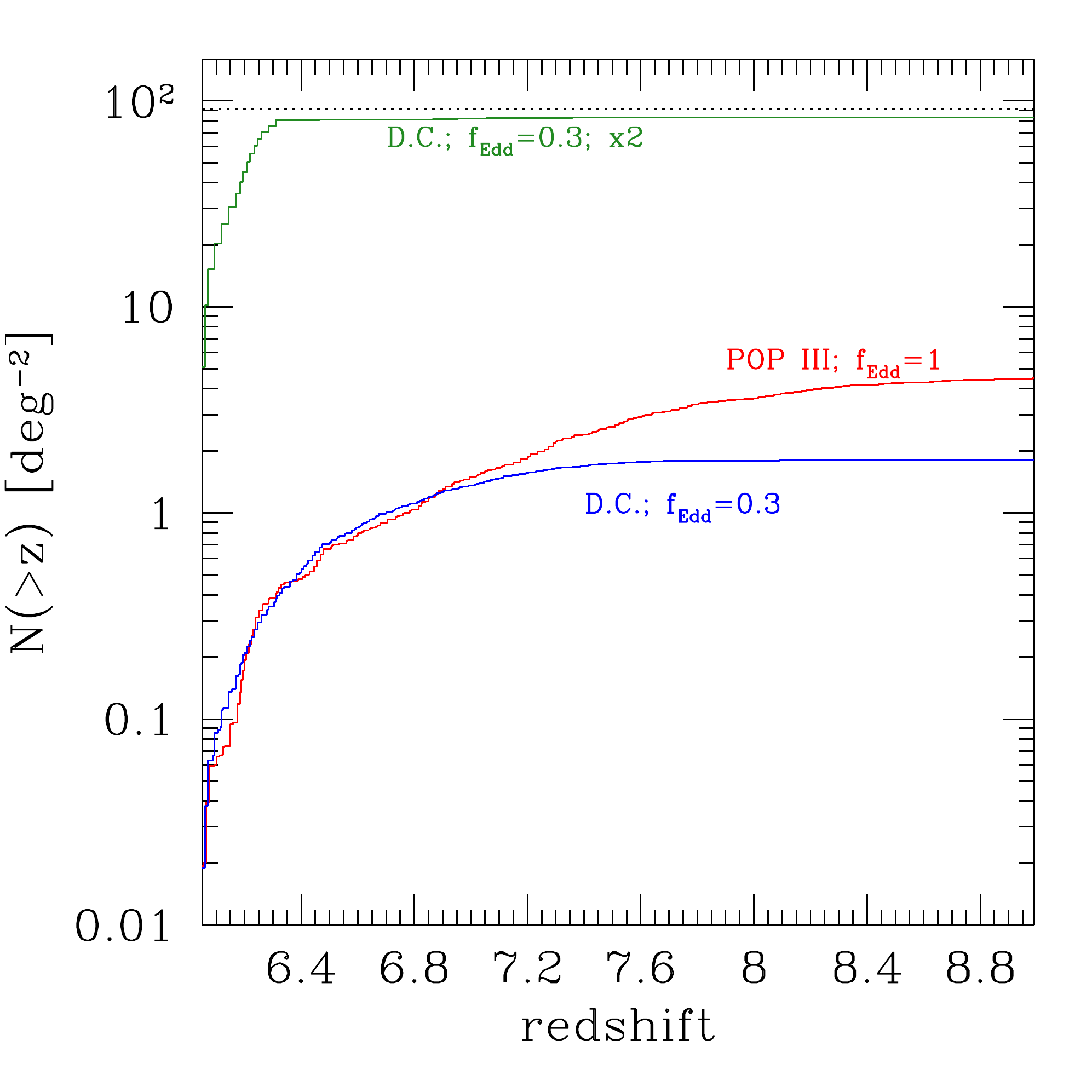}
\end{center}
\caption{Cumulative number of sources as a function of redshift for individual X-ray detections. This calculation assumes the X-ray flux limit of the 4 Msec CDF-S {\it Chandra} 
observations. The {\it horizontal dotted line} shows the number density required to individually detect one source in the area considered in this work at $z>$7. Models 
are described in the supplemental material and  labelled in the figure (Pop III, $f_{\rm Edd}=1$; D.C., $f_{\rm Edd}=0.3$; D.C., $f_{\rm Edd}=0.3$, $\times 2$). Note:  model 
Pop III, $f_{\rm Edd}=1$, $\times 2$ has no detectable source. To distinguish between these models for early black hole formation will require a deep 
multiwavelength survey covering at least $\sim$1~deg$^{2}$.}
\end{figure}

\clearpage

\begin{center}
\title  {\huge \bf{Supplementary Information}}
\end{center}
\author{}
\date{}


\section*{The Observations\\}

\begin{figure}
\includegraphics[scale=0.45]{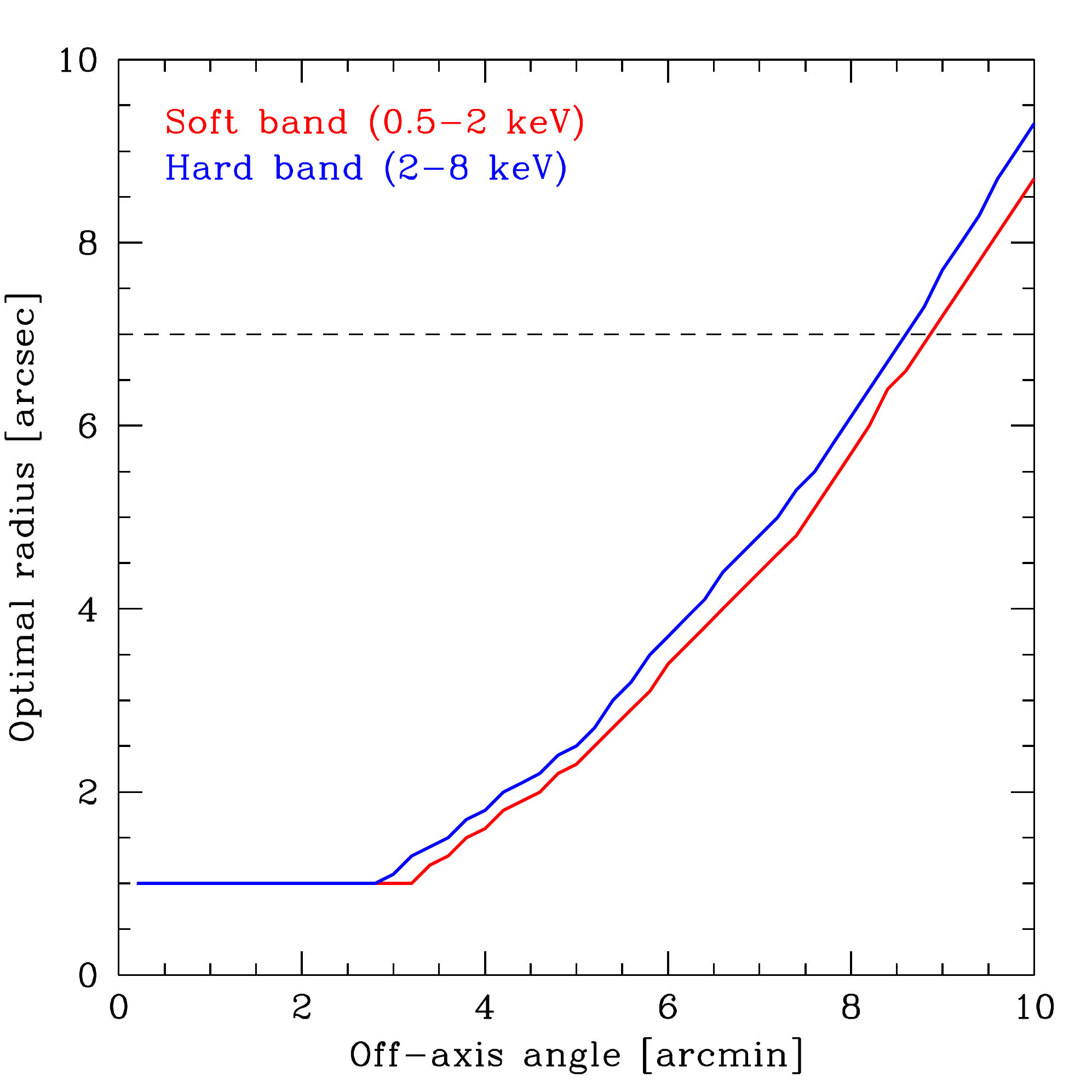}
\caption{Optimal extraction radius as a function of off-axis angle, measured from the average {\it Chandra} pointing
center. As described in the appendix A, these lines were measured by optimizing the function $f(\theta,r)$/$r$. A minimum radius of 1$''$ was assumed
in order to avoid flux losses due to astrometric problems and pixel aliasing. {\it Red} 
and {\it blue} lines show the optimal radii for the soft and hard band respectively. Sources with optimal radii greater
than 7$''$ ({\it dashed horizontal line}) are not considered in the stack, given their low expected contribution to the
integrated signal-to-noise.}
\label{opt_rad_angle}
\end{figure}

The main observational data samples for this work are the 371 galaxy candidates at
$z$$\sim$6 selected using the optical and near-IR Lyman break technique\cite{bouwens06}, together with
66 $z$$\sim$7 and 47 $z$$\sim$8 galaxy candidates\cite{bouwens10}, all of them in the {\it Chandra} Deep 
Field South (CDF-S) field. We then complemented this sample with the 151 $z$$\sim$6 candidates in the {\it Chandra} Deep 
Field North (CDF-N)\cite{bouwens06}. The accuracy of this drop-out technique selection has been recently confirmed by spectroscopic observations of 
some of these sources\cite{vanzella09,vanzella10,lehnert10}. The contamination by foreground sources in these samples appears to be 
very small, $\sim$10\%\cite{vanzella09,bouwens10}. We then used  the 4 Msec {\it Chandra} observations of the CDF-S\cite{note1}  
and the 2 Msec {\it Chandra} data available on the CDF-N\cite{alexander03} in
order to search for signatures of supermassive black hole (SMBH) accretion in these sources.  None of these galaxies are 
detected individually in the X-ray data. However, {\it Chandra} data are uniquely suited to perform
stacking, which allows us to reach much fainter flux limits. In order to maximize the signal-to-noise of these measurements
we used an optimized X-ray stacking scheme, described in detail in Appendix A. We stack independently in the 
soft, 0.5-2 keV, and hard, 2-8 keV, observed-frame {\it Chandra} bands. At $z$$\sim$7, they correspond to rest-frame 
energies of 4-16 and 16-64 keV respectively. While at these high rest-frame energies the effects of Compton-thin 
obscuration ($N_H$$<$10$^{24}$cm$^{-2}$) are negligible, we have to consider the possibility of higher absorption column 
densities. We restricted our stack to sources closer than $\sim$9$'$ from
the average aim point of the {\it Chandra} observations in order to have an optimal extraction radius smaller than
7$''$ (Fig.~S\ref{opt_rad_angle}). We further removed sources with an X-ray detection closer than 22$''$ to avoid
possible contamination in the background determination.

\begin{figure}
\begin{center}
\includegraphics[scale=0.25]{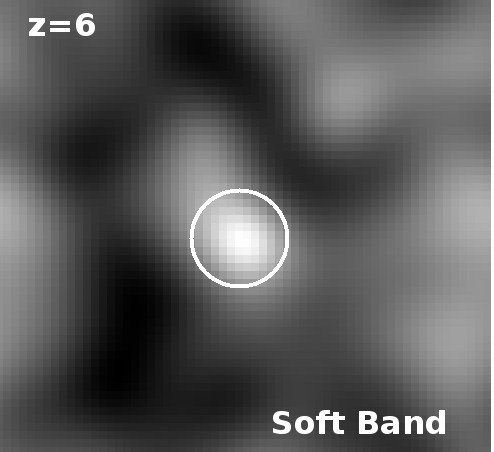}
\includegraphics[scale=0.25]{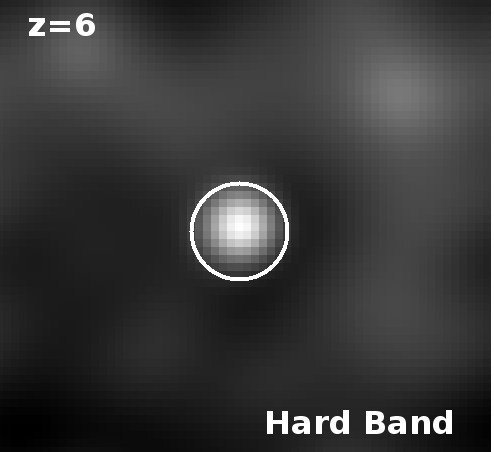}
\end{center}
\caption{Stacked {\it Chandra} images for the $z$=6 galaxy sample in the soft ({\it left panel}) and
hard ({\it right panel}) X-ray bands. The detections are significant at the 5 and 6.8 -$\sigma$ levels
respectively. Each image is 30$''$$\times$30$''$. The {\it white circle} at the center of each image has a radius
of 3$''$. Images were adaptively smoothed using a minimum scale of 3 pixels, a maximum scale of 5 pixels  and 
minimum and maximum significances of 3 and 6 respectively.}
\label{z6_stack}
\end{figure}

In the sample of candidates at $z$$\sim$6 we stacked a total of 197
sources, 151 in the CDF-S and 46 in the CDF-N. This corresponds
to a total exposure time of $\sim$7$\times$10$^8$ seconds ($\sim$23
years). We found significant detections, $\geq$5-$\sigma$, in both the 
soft and hard bands independently. In the soft band we computed a count
rate of 3.4$\pm$0.68$\times$10$^{-7}$ counts~s$^{-1}$ per source, which 
corresponds to an observed-frame soft band flux of 2.3$\times$10$^{-18}$~erg~cm$^{-2}$s$^{-1}$.
Converting it to the rest-frame hard band we obtain a flux 
of 1.9$\times$10$^{-18}$~erg~cm$^{-2}$s$^{-1}$. In the observed-frame hard
band we measure a count rate of 8.8$\pm$1.3$\times$10$^{-7}$ counts~s$^{-1}$, 
which corresponds to a 6.8-$\sigma$ detection. The average flux in the observed-frame
hard band is 2.1$\times$10$^{-17}$~erg~cm$^{-2}$s$^{-1}$, which converted to
the rest-frame hard-band corresponds to 1.7$\times$10$^{-17}$~erg~cm$^{-2}$s$^{-1}$.
Stacked {\it Chandra} images for this sample in the soft and hard X-ray bands are shown in Fig.~S\ref{z6_stack}

\begin{figure}
\begin{center}
\includegraphics[scale=0.45]{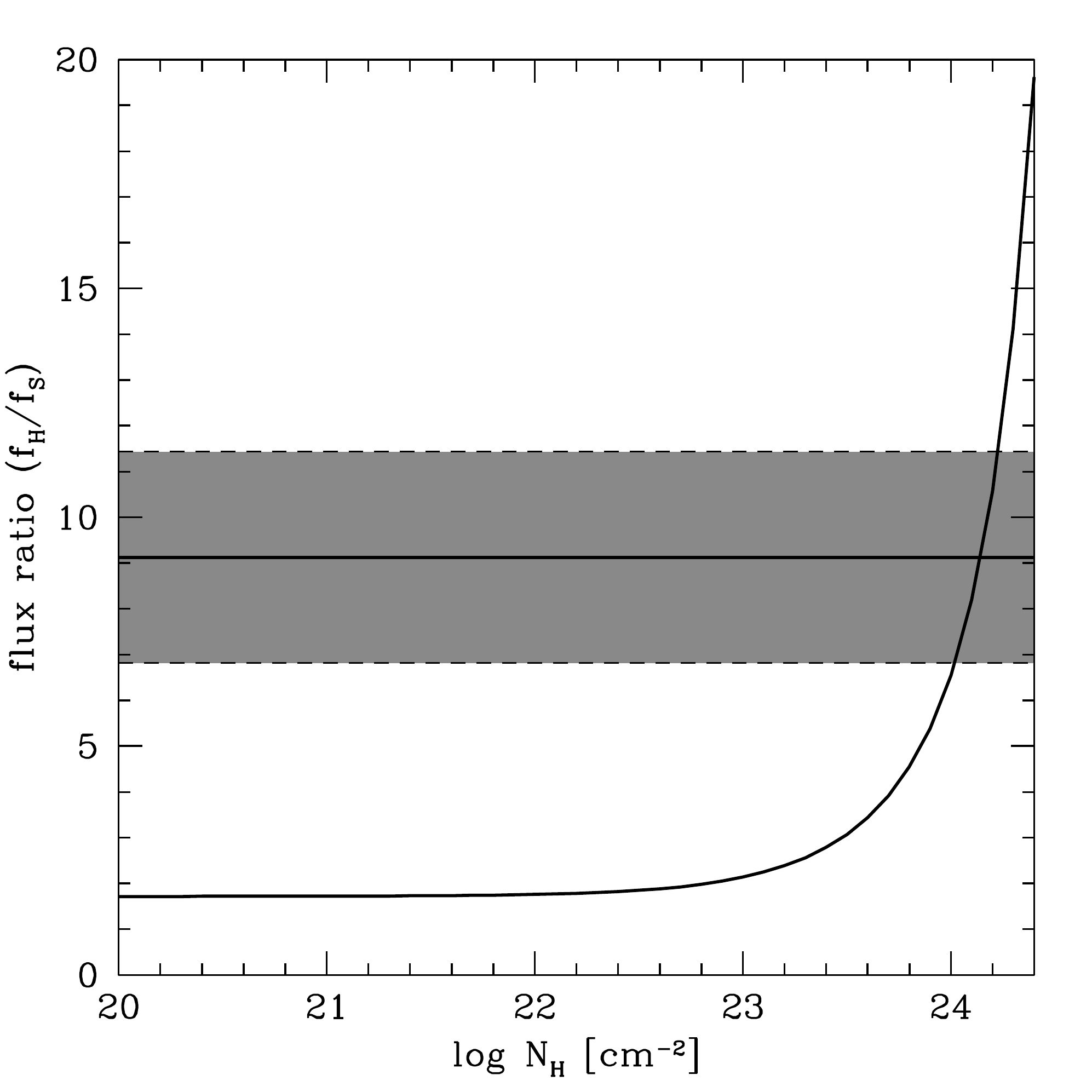}
\end{center}
\caption{Expected ratio of the observed-frame hard to soft flux as a function of obscuring neutral Hydrogen column density ($N_H$).
The {\it black solid line} was derived assuming an intrinsic power-law spectrum with slope $\Gamma$=1.9 and photoelectric absorption.
The gray zone shows the measured ratio for the stack of galaxies at $z$$\simeq$6 and the $\pm$1 s.d. limits. A column density of
N$_H$$\simeq$10$^{24}$~cm$^{-2}$, i.e. Compton-thick obscuration, is required to explain the observed hard to soft X-ray flux ratio.}
\label{fratio_nh}
\end{figure}

There is a factor $\sim$9 difference between the fluxes measured in the
observed-frame soft and hard bands. Assuming a power-law X-ray
spectrum with an intrinsic spectral slope $\Gamma$=1.9, typical of AGN\cite{nandra97}, the 
expected flux ratio between the observed-frame soft and hard bands for an unobscured
source is expected to be $\sim$1.7 (Fig.~S\ref{fratio_nh}). The only explanation for the relatively 
large flux ratio in the hard to soft bands is very high levels of obscuration. As can be seen in Fig.~S\ref{fratio_nh},
at $z$$\sim$6 a minimum column density of N$_H$$\simeq$10$^{24}$~cm$^{-2}$, i.e. Compton-thick 
obscuration, is required. Given that this ratio is observed in the stack, this implies that
there are very few sources with significantly lower levels of obscuration, which in turn means
that these sources must be nearly Compton-thick along most directions ($\sim$4$\pi$ obscuration). Similar sources 
have also been observed in the local Universe\cite{ueda07} but are likely rare. Furthermore, up to $z$$\sim$3
it has been shown before\cite{ueda03,lafranca05,sazonov07} that the fraction of obscured AGN increases with
decreasing luminosity and increasing redshift\cite{treister06b,ballantyne06}. Hence, it is not entirely surprising that the 
sources studied here, given their low luminosities and high redshifts, are heavily obscured. In fact, the discovery of a 
Compton-thick AGN at $z$$\sim$5 selected using the drop-out technique has been recently reported\cite{gilli11}.

The corresponding average rest-frame 2-10 keV luminosity, derived from the observed-frame hard 
band, is 6.8$\times$10$^{42}$~erg~s$^{-1}$. Since none of these sources are individually detected in X-rays,
we conclude that at least 30\% of the galaxies in this sample contain an AGN. Multiplying by the number of sources we obtain a total 
luminosity of 1.34$\times$10$^{45}$~erg~s$^{-1}$. The total area surveyed is $\sim$310 arcmin$^2$, or 
0.086 deg$^2$. Hence, the integrated AGN emissivity at $z$$\sim$6 derived from this sample is
1.6$\times$10$^{46}$~erg~s$^{-1}$deg$^{-2}$.

Similarly, we stacked a total of 57 sources by combining the samples 
at $z$$>$7 and $z$$>$8. For the galaxies in this redshift range, we found
no significant detection in the observed-frame soft or hard
bands. The 3-$\sigma$ upper limits are 6.9$\times$10$^{-7}$ cts s$^{-1}$
and 1.4$\times$10$^{-6}$ cts s$^{-1}$, which corresponds to flux upper limits of
4.6$\times$10$^{-18}$~erg~cm$^{-2}$~s$^{-1}$ and
3.3$\times$10$^{-17}$~erg~cm$^{-2}$~s$^{-1}$ respectively.  Assuming an average
redshift $z$=7.5 and converting to the rest-frame hard band, 2-10 keV,
we obtain average luminosities of 3.1$\times$10$^{42}$~erg~s$^{-1}$ and
2.2$\times$10$^{43}$~erg~s$^{-1}$ from the observed-frame soft and
hard band respectively. Because we only have upper limits in both bands, we 
cannot constrain the presence of obscuration in this sample, however assuming that these
AGN are as heavily obscured as the $z$$\sim$6 sample, the shallower limit from the
hard band is actually the more stringent constraint. The area covered in the
$z$$>$7 observations is 40 arcmin$^2$, or 0.011 deg$^2$. Hence, the
upper limits to the integrated AGN emissivity are
5.2$\times$10$^{45}$~erg~s$^{-1}$deg$^{-2}$ and
4.3$\times$10$^{46}$~erg~s$^{-1}$deg$^{-2}$ respectively.

Finally, we also stacked the galaxies in the $z$$\sim$7 sample
separately. In this case we obtained 3-$\sigma$
upper limits of 6.9$\times$10$^{-18}$~erg~cm$^{-2}$s$^{-1}$ and
5.0$\times$10$^{-17}$~erg~cm$^{-2}$s$^{-1}$ in the observed frame soft and hard
bands respectively. These correspond to average observed-frame 
luminosities of 4.0$\times$10$^{42}$~erg~s$^{-1}$ and
2.9$\times$10$^{43}$~erg~s$^{-1}$. Dividing by the survey area and
multiplying by the number of sources we obtain 3-$\sigma$ upper limits
for the integrated AGN emissivities from these galaxies of
4.3$\times$10$^{45}$~erg~s$^{-1}$deg$^{-2}$ and
2.9$\times$10$^{46}$~erg~s$^{-1}$ respectively.

\section*{Black Hole Mass Density Determination\\}

We compute the observed integrated black hole mass density using
an updated version of the ``Soltan''\cite{soltan82} argument. Following
the standard derivations\cite{yu02,marconi04} we have that the
integrated black hole mass density is given by:

\begin{equation}
\rho_{BH}(z)=\int^\infty_z\frac{dt}{dz}{dz}\int^\infty_0\frac{1-\epsilon}{\epsilon c^2}L_{bol}\Psi(L,z)dL,
\end{equation}

where

\begin{equation}
L_{bol}=k_{corr} L_X
\end{equation}

and $k_{corr}$ is the bolometric correction for the rest-frame hard
X-ray band. In order to compute $\rho_{BH}(z)$ we need to make some
assumptions, since the AGN luminosity function (LF) at $z$$>$6 is unknown, and only the 
integrated luminosity is known. First, we assume that at $z$$>$6 the AGN LF does not 
depend strongly on redshift, i.e., $\Psi (L,z)$$\simeq$$\Psi (L)$. We further assume that 
the efficiency and bolometric corrections are constant (i.e., independent of luminosity). While 
we know that the bolometric correction depends on luminosity\cite{marconi04,barger05}, our 
sample most likely does not span a wide luminosity range, and hence this assumption is
reasonable. Therefore, we have,

\begin{equation}
\rho_{BH}(z)=\frac{(1-\epsilon)k_{\rm corr}}{\epsilon c^2}\int^\infty_z\frac{dt}{dz}{dz}\int^\infty_0L_X\Psi(L)dL.
\end{equation}

The second integral on the right hand side can be determined from the
observed integrated AGN emissivity. We only need to convert from a
density per unit of sky area (the observed value) to a co-moving AGN
emissivity per unit of volume. This can be done easily by dividing the
observed value by the co-moving volume at the redshift range covered by each sample. 
In principle, a correction should be made to account for the contribution of high-luminosity
sources not present in the relatively narrow fields considered in this work. However, as we will 
show below, this correction should be small, $\sim$1-2\%, and thus can be safely ignored.

The bolometric correction $k_{\rm corr}$ has been estimated by several authors in the past.
For high-luminosity sources (quasars), a value of 35 was measured\cite{elvis94}. For 
low-luminosity sources, L$_x$$\sim$10$^{43}$ erg~s$^{-1}$, values of $\sim$10-20, were estimated\cite{marconi04},
while newer calculations\cite{natarajan09}  report a value of $\sim$25 for sources with  L$_x$$\sim$10$^{42}$ erg~s$^{-1}$ and $\sim$40 at 
L$_x$$\sim$10$^{43}$ erg~s$^{-1}$. Given the low luminosity of the sources in our sample we assume a value of $k_{\rm corr}$=25, 
with an estimated uncertainty of a factor of $\sim$2. The main factor contributing to this uncertainty is the determination of the contribution of the 
X-ray emission relative to the ultraviolet (where most of the bolometric output is found). Observational studies show that  this factor depends 
on luminosity but remains constant with redshift up to $z$$\sim$6\cite{steffen06}.

Assuming a constant radiation efficiency $\epsilon$=0.1 we obtain from the combined $z$=7-8 sample 
3-$\sigma$ upper limits of $\rho_{BH}$$<$708$~M_\odot$Mpc$^{-3}$ (possibly affected by obscuration)
and $\rho_{BH}$$<$5883$M_\odot$Mpc$^{-3}$ from the observed-frame soft and
hard band respectively.

For the sample of galaxies at $z$$\sim$7 only, assuming that all these galaxies 
lie between $z_1$=7 and $z_2$=8, we obtain that $\rho_{BH}$$<$1117~$M_\odot$ Mpc$^{-3}$
from the observed-frame soft-band and $\rho_{BH}$$<$7595~$M_\odot$ Mpc$^{-3}$
from the hard band. Finally, for our sample of $z$$\sim$6 galaxy candidates
we obtain from the stacked detection in the observed-frame hard band
that $\rho_{BH}$=5005$\pm$751 $M_\odot$ Mpc$^{-3}$ (1-$\sigma$).

\begin{figure}
\begin{center}
\includegraphics[scale=0.45]{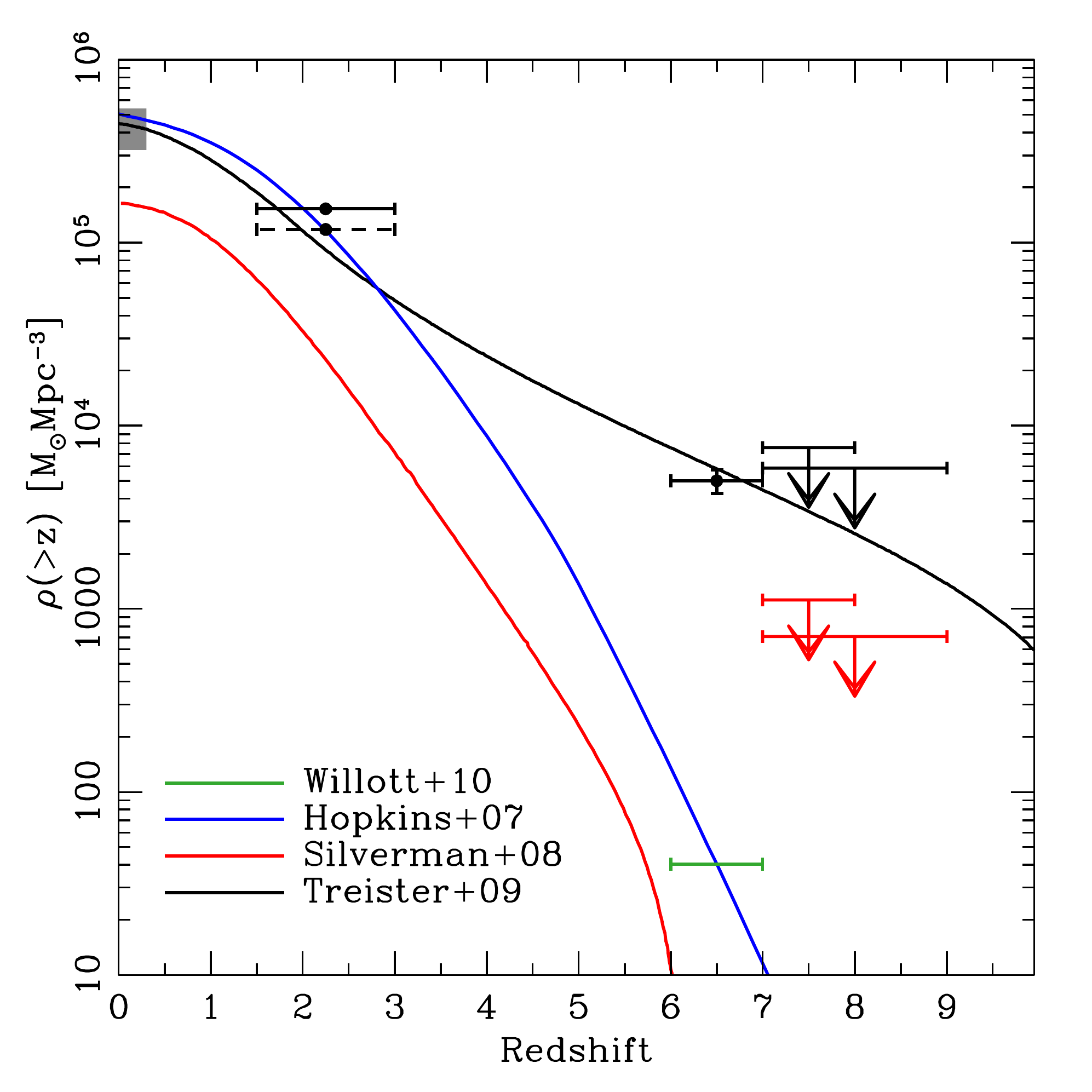}
\end{center}
\caption{Accreted black hole mass density as a function of redshift. The {\it gray rectangle} shows the 
range of values allowed by observations of $z$$\simeq$0 galaxies\cite{shankar09}. The data points at $z$$\sim$2 correspond
to the values obtained from {\it Chandra} observations of X-ray detected AGN and luminous infrared galaxies\cite{treister10b}, while 
the measurements at $z$$\sim$6 and the upper limits at $z$=7-9 show the results described in this work ({\it red} and {\it black} data points from the observed-frame soft and hard X-ray band observations respectively). The {\it red} and {\it blue}  lines show the 
values inferred from AGN luminosity functions\cite{silverman08b,hopkins07}, while the point at $z$=6-7 was obtained from the quasar 
luminosity function\cite{willott10a}. The {\it black line} assumes the hard X-ray AGN LF of Ueda et al.\cite{ueda03}, as modified by Treister et al.\cite{treister09b}.}
\label{bhmass_w_z}
\end{figure}

In Figure~S\ref{bhmass_w_z} we plot the integrated accreted black hole
mass density as a function of redshift. The upper limits at $z$$\sim$7-8 and
measurement at $z$$\sim$6 obtained from our work are shown together with the observations in the
local Universe\cite{shankar09} and the values derived from {\it Chandra} observations of 
X-ray detected AGN and luminous infrared galaxies at $z$$\sim$2\cite{treister10b}.

\section*{Accretion Models: key features\\}
We investigate the formation and evolution of black holes via cosmological realizations of the 
merger hierarchy of dark matter halos from early times to the present in a $\Lambda$CDM cosmology.
The main features of the by now fairly standard SMBH evolution models that are used to interpret the 
data have been discussed in detail elsewhere\cite{VHM,VN09,Volonteri2010}. 
We briefly outline some of the key features here. In this work,  two ``seed'' formation 
models are considered : those deriving from population-III star remnants (Pop III), and from direct collapse models (D.C.).
The main difference between these two models lies in the mass function of 
seeds. Pop III seeds are light weight (few hundred solar masses) 
and form abundantly and early (roughly one per comoving cubic Mpc at $z\simeq20$). 
D.C. seeds are more massive ($10^4-10^6$ solar masses), but rarer 
(a peak density of 0.1 per comoving cubic Mpc at $z\simeq12$).
We summarize below  the relevant and standard assumptions that go into our 
modeling of SMBH growth.

Essentially in this scheme central SMBHs hosted in galaxies accumulate mass via accretion 
episodes that are triggered by galaxy mergers.  Accretion proceeds in one of two modes:  self-regulated or un-regulated.  
For each SMBH in our models we know its mass at the time when the merger starts ($M_{\rm in}$), 
and we set the final mass through the self-regulated or un-regulated prescription.
These two models differ by the amount of mass a SMBH accretes during a given accretion phase. 
In the context of the currently supported paradigm for
structure formation, growth of structure in the Universe occurs hierarchically and via copious merging 
activity. Our models rely on the following assumptions:

$\bullet$ SMBHs in galaxies undergoing a major merger (i.e., having a mass ratio
$>$1:10) accrete mass and become active. 

$\bullet$ In the self-regulated model, each SMBH accretes an amount of mass, corresponding to 90\% of the
mass predicted by the local $M_{\rm BH}-\sigma$ relation\cite{Gultekin2009}, 
\beq
M_{\rm fin}=M_{\rm in}+ 0.9 \times 1.3\times10^8  \left(\frac{\sigma}{200 \kms} \right)^{4.24} \msun;
\eeq
the 90\% normalization was chosen to take into account the contribution of mergers, without largely exceeding the mass given by the $M_{\rm BH}-\sigma$ relation for SMBHs at $z=0$. Here $\sigma$ is the velocity dispersion of the host after the merger.  
We adopt  $\sigma=V_c/\sqrt[]{3}$, where $V_c$ is the virial velocity of the host dark matter halo\cite{Ferrarese2002, Kormendy2004}. A SMBH is assumed to stop accreting once it reaches the value given by the  $M_{\rm BH}-\sigma$ relation. 

$\bullet$In the unregulated  mode ($\times 2$) we simply set $M_{\rm fin}=2\times M_{\rm in}$, that is, we 
double the mass during each accretion episode. 

$\bullet$ The  rate at which mass is accreted scales with the Eddington rate ($f_{\rm Edd}$) for the SMBH, where $f_{\rm Edd}$ is the 
accretion rate in units of the Eddington rate.  We adopt   $f_{\rm Edd}=0.3$ 
for D.C. models and $f_{\rm Edd}=1$ for Pop III models in order to reproduce the mass
density at $z=0$. As Pop III seeds are lighter, they need to accrete more mass over the 
course of their cosmic history to become supermassive.

$\bullet$  The lifetime of an AGN depends on how much mass it accretes during each episode. Given the initial mass of a SMBH, $M_{\rm in}$, and the amount of mass it accretes, we can calculate its final mass, $M_{\rm fin}$, at the end of the active phase (Equation 4). For a given Eddington fraction, $f_{\rm Edd}$, the mass of the SMBH grows with time as:
\beq
M_{\rm fin}=M_{\rm in} \exp\left(f_{\rm Edd}\frac{1-\epsilon}{\epsilon}\frac{t}{t_{\rm Edd}}\right)
\eeq
where $\epsilon$ is the radiative efficiency ($\epsilon \simeq 0.1$), $t_{\rm Edd}=M_{\rm BH} c^2/L_{\rm Edd}=\frac{\sigma_T \,c}{4\pi \,G\,m_p}\simeq 0.45$ Gyr ($c$ is the speed of light, $\sigma_T$ is the Thomson cross section, $m_p$ is the proton mass). Given $M_{\rm in}$ and $M_{\rm fin}$, then a SMBH is active for a time $t_{\rm AGN}=\frac{t_{\rm Edd}}{f_{\rm Edd}} \frac{\epsilon}{1-\epsilon}\ln(M_{\rm fin}/M_{\rm in})$.  Note that the un-regulated prescription (a SMBH mass doubles at each accretion episode) corresponds to assuming that all SMBHs shine as AGN for the same time at a fixed accretion rate (e.g., 100 Myr for an accretion rate $f_{\rm Edd}=0.3$).

In summary, we study and compare two self-regulated models (Pop III, $f_{\rm
Edd}=1$; D.C., $f_{\rm Edd}=0.3$) and two unregulated models
(Pop III, $f_{\rm Edd}=1$, $\times 2$; D.C., $f_{\rm Edd}=0.3$,
$\times 2$).  We use these models to derive the mass density in black holes in a cosmic volume, by summing over
all existing black holes at a given redshift and normalizing by the comoving volume. 
We also calculate number counts (Figure 2 in the main article) imposing selection criteria that match the observations
we have analyzed.  The limiting luminosity is $4.2\times 10^{41}$~erg~s$^{-1}$ for the stacked sample, i.e., if all the galaxies are emitting
at this limit then we should detect them in the stack. For a single source, the more appropriate
limit is this value multiplied by the number of sources in the stack, which is equivalent to a single
source contributing all the signal. In that case, the luminosity limit is $3.1\times 10^{43}$~erg~s$^{-1}$
and the flux limit is  $4.95\times 10^{-17}$~erg~cm$^{-2}$~s$^{-1}$. In that case, the source will be individually detected. We therefore 
select all accreting black holes in the theoretical sample at $z>6$ that meet this flux criterion. From the Eddington ratio we calculate the 
bolometric luminosity and then apply a K-correction of 25 to derive the X-ray flux at the appropriate redshift.

\section*{Comparison with AGN Luminosity Functions\\}

We first compare the observational results presented with the expectations 
from integrating existing measurements or extrapolations of the observed AGN luminosity 
functions from earlier work. Integrating the hard X-ray LF\cite{ueda03} from $z$=7 to 10 we obtain 
an integrated luminosity density of 2.3$\times$10$^{46}$~erg~s$^{-1}$deg$^{-2}$, in
good agreement with the values obtained from the stacking in the hard
X-ray band, which are less affected by obscuration. In contrast, from 
the LF of $z$$\sim$6 quasars\cite{willott10a}, we expect a steep 
decline in the AGN number density. 

Comparison of observations of the accreted SMBH mass density with 
predictions obtained from extrapolations of existing AGN luminosity functions to
high redshifts and/or low luminosities provides contrasting
results. While there is in general good agreement up to $z$$\sim$3,
differences of up to $\sim$2 orders of magnitude are found at
$z$$>$6. This can be explained as the luminosity functions of
Hopkins et al.\cite{hopkins07} and Willott et al.\cite{willott10a} incorporate a density
evolution of the form 10$^{-0.47z}$, which is required to account for the observed
evolution of the highest luminosity quasars at $z$$\sim$5-6 in the
SDSS\cite{fan01}. A similarly steep evolution from $z$$\sim$3 to
$z$$\sim$5 has been reported, based on a compilation of {\it Chandra} X-ray observations\cite{silverman08b}. However, it 
is worth pointing out that the accreted SMBH mass density inferred from this LF,
presented in Figure~S\ref{bhmass_w_z} is significantly lower than
observations and other LFs at all redshifts, which may be explained by
the relatively high spectroscopic incompleteness ($\sim$50\%) of this sample. Incompleteness
is a particularly important issue and limitation at high redshift, where sources are faint in
the optical bands. In contrast, the AGN LF of Treister et al.\cite{treister09b} is in good agreement 
with the observations at all redshifts, and it is therefore the only one plotted in Figure 1 of 
the main paper. This work is based on the Ueda et al.\cite{ueda03} hard X-ray AGN LF modified
to incorporate an increasing number of obscured sources with increasing redshift\cite{treister06b}
and reducing the relative number of Compton-thick sources by a factor of $\sim$4, consistent with the
observed space density of these sources in the all-sky {\it INTEGRAL} and {\it Swift} surveys.

Explaining the accreted SMBH density at $z$$>$6 inferred from 
X-ray stacking observations would require that the comoving space
density of low-luminosity AGN, $L_X$$<$10$^{44}$~erg~s$^{-1}$, remains
nearly constant up to $z$$\sim$3\cite{ueda03}.  At
the same time, the discrepancy with the observed density of
high-luminosity sources in optical surveys can be resolved if there
is a strong evolution in the number of high-luminosity sources. For example,
a decline in the number of quasars with redshift given by 10$^{-0.47z}$ has been 
measured\cite{willott10a}. In this case, the contribution from high-luminosity
sources to the integrated accreted black hole mass density will be very small,
$\sim$1-2\%. Alternatively, the observed results could be explained if the
relative number of heavily-obscured quasars increases strongly with
redshift, as measured up to $z$$\simeq$3\cite{treister10}. 

\section*{Implications for re-ionization\\}

Two sources of UV radiation are expected to contribute to re-ionization: star forming galaxies and quasars\cite{Shapiro87,Haardt96,Madau99}. 
The conventional view is that galaxies dominate hydrogen re-ionization occurring at high redshift\cite{spergel07}, $z\simeq 9$,
and quasars dominate helium re-ionization occurring at a later time, $z\simeq 3$\cite{Madau99}.  This result is based 
on optically-selected quasar luminosity functions\cite{hopkins07,willott10a} that show, as discussed in the previous section,
a steep drop of the quasar population at $z$$>$4\cite{Faucher2008}. However, in ab initio models of black 
hole evolution through cosmic history\cite{volonteri09} the amount of black hole growth required to explain the bright end of the 
luminosity function of quasars at $z=3-6$ implies a substantial contribution of quasars to hydrogen re-ionization, provided that UV and X-ray 
radiation can escape into the intergalactic medium.  Our findings resolve this tension, as regardless of the amount of accretion occurring onto black 
holes, the bulk of the emission from the population is obscured, and UV photons cannot escape. In 4$\pi$ Compton thick sources only the 
highest energy photons can escape.  We provide a simple estimate here of the contribution of Compton thick quasars to re-ionization. From our 
models we can extract information on $d\rho$, the accreted mass density on black holes as a function of cosmic time (Figure 1 in the main article shows the integral 
quantity). Let us assume, optimistically,  that an escape fraction $f_{esc}=0.04$ of the bolometric luminosity (here we just assume $f_{esc}=1/k_{corr}$) 
goes into hard X-ray photons with mean energy $E_\gamma=10$ keV that can escape from the quasar host. All these photons contribute to primary and 
secondary ionizations. The number of ionizing photons per hydrogen atom can be evaluated from the following differential equation:
\begin{equation}
d x=\frac{f_{esc}}{\rho_{H}E_\gamma} d \rho+
\frac{f_{\rm  SI}(x)}{\rho_{H}14.4 \rm{eV} } d \rho
  -x\frac{dt}{t_{rec}},
\end{equation}
where $\rho_{H}$ is the hydrogen cosmic density\cite{spergel07}, and the recombination timescale for hydrogen is $t_{rec}\simeq0.3[(1+z)/4]^{-3}$ Gyr\cite{Madau99}. Here  
the first term of the right hand side of the equation includes primary ionizations by ionizing photons, the second term accounts for secondary
ionizations from energetic ionizing photons \cite{ss85,Madau2004,volonteri09}, and the third term accounts for recombinations.
Here $  f_{\rm SI}(x)\approx0.35(1-x^{0.4})^{1.8} - 1.77\left(\frac{28\rm{eV}}{E_\gamma}\right)^{0.4}x^{0.2}(1-x^{0.4})^2$, as long as $x<1$, and $x$ is less than unity  in all our models down to $z\simeq 1$.

\begin{figure}[h!]
\begin{center}
\includegraphics[scale=0.45]{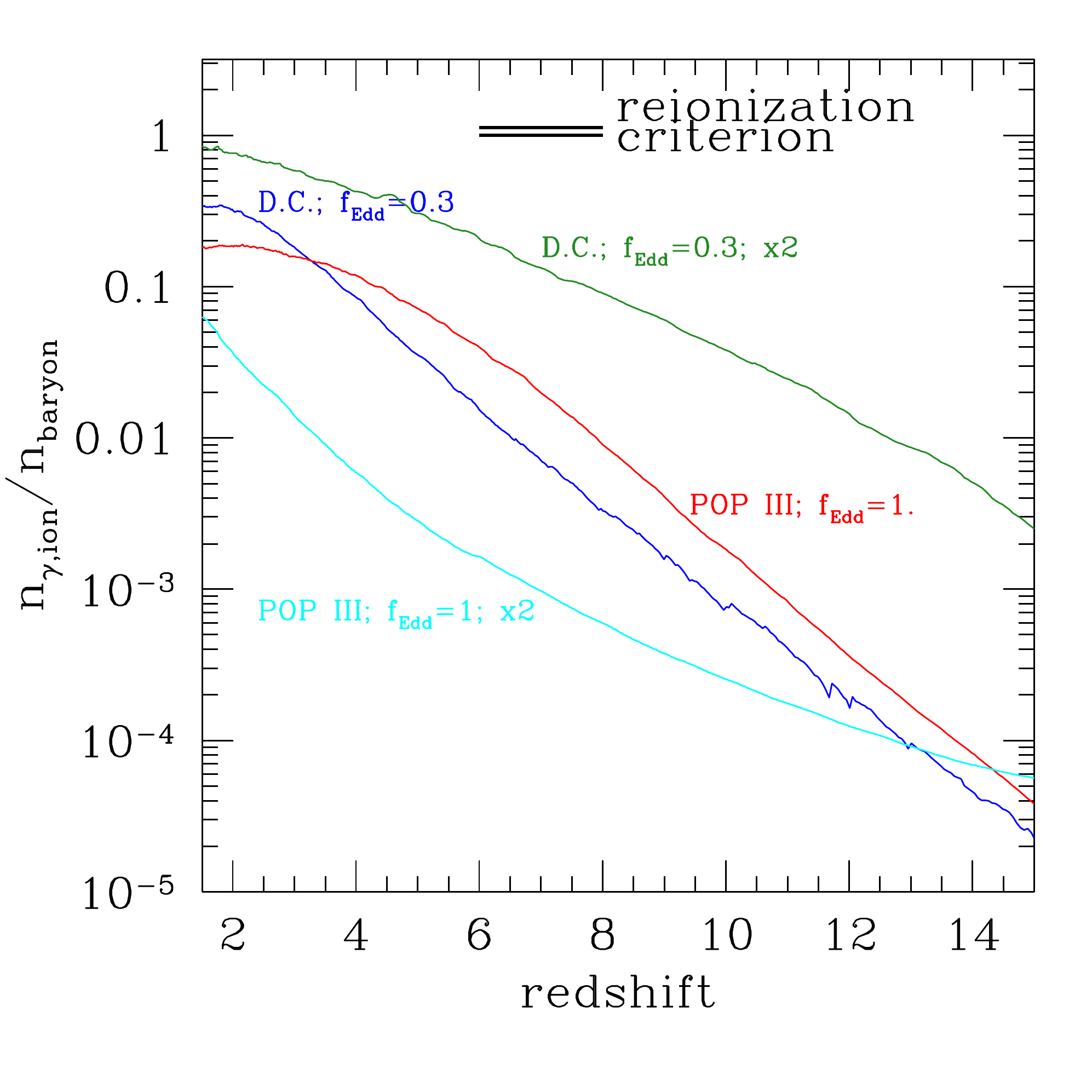}
\end{center}
\caption{Number of ionizing photons per hydrogen atom. Models  are labelled in the figure (Pop III, $f_{\rm Edd}=1$; Pop III, $f_{\rm Edd}=1$, $\times 2$; D.C., $f_{\rm Edd}=0.3$; D.C., $f_{\rm Edd}=0.3$, $\times 2$). A ratio of at least one ionizing photon per hydrogen atom is required to re-ionize the Universe. As can be seen in this figure,
most models predict a ratio for the early growing black holes that is lower than this value by 2-3 orders of magnitude at $z$$>$6.}
\label{reion_ct}
\end{figure}

The  number of ionizing photons per hydrogen atom is shown in the accompanying Figure~S\ref{reion_ct}, where we see that models that successfully reproduce 
our observations {\bf fail} to reionize hydrogen by 2-3 orders of magnitude at $z>$6, notwithstanding significant black hole growth. Without taking obscuration into 
account, these accreting black holes would contribute substantially to re-ionization, producing one ionizing photon per atom by $z=6$\cite{volonteri09}. 

However, we stress here that if quasars do not contribute to hydrogen re-ionization, it is not because quasar activity drops steeply 
at $z$$>$4 as previously suggested\cite{Faucher2008}, but instead due to the presence of obscuring material that absorbs UV and soft X-ray radiation. 
As seen in Figure~S\ref{reion_ct}, the models that are consistent with the observations at $z$$>$6 and include self-regulation under-produce ionizing photons 
at these epochs. Hence, re-ionization of the early Universe is most likely due to star forming galaxies and not growing black holes. However, this is still vigourously 
debated. As previously shown\cite{ouchi09}, observed $z$$\sim$7 galaxies cannot provide enough hydrogen-ionizing photons unless some of the galaxy properties, such 
as escape fraction, metallicity, initial stellar mass function or dust extinction, evolve significantly from $z$$\sim$7 to the local Universe. Similarly, the extrapolation of the 
galaxy luminosity function at the faint end plays a major role in whether these sources can re-ionize the Universe\cite{grazian10}.

\newpage

\section*{Appendix A: Optimized X-ray stacking\\}

We start by assuming that all sources are equally bright in the X-rays, 
with count rate $R_s$, $R_h$ in cts/Ms in the soft and hard bands, respectively.
The background levels in {\it Chandra} for the ACIS-I detectors as measured in the
CDF-S 4 Msec data are $B_s$=0.049 cts/Ms/pixel in the soft band and $B_h$=0.160 cts/Ms/pixel
in the hard one, in good agreement with the values of 0.066 cts/Ms/pixel (soft), and 0.167 cts/Ms/pixel (hard)
measured  in the CDF-S 2 Msec data\cite{luo08}. The background level was found to vary by $\sim$5\% 
(standard deviation) across the field. To account for this, the background value was determined for each source independently
by measuring the average count rate on a annulus around the source position with inner radius 2$r$, where $r$ is the aperture
radius in pixels and outer radius fixed to 22$''$. A 3-$\sigma$ clipping was used for the background determination. Then, an aperture 
of radius $r$ pixels has area $\pi$$r^2$ and hence background counts $B_k$$\pi$$r^2$$<E(r)>_k$ where $<E(r)>_k$ stands for the 
average of the exposure time $E_{ij}$ over all pixels inside this aperture around the kth source and $B_k$ is the background level
for that source.  The signal contained within this aperture can be found by interpolating the enclosed energy profiles, which give us the 
values of $r$ corresponding to various fractions $f$ of $R$ as a function of offaxis angle $\theta$, e.g. $r(\theta,f=0.95)$. We interpolate 
between these curves to determine the desired function, $f(\theta,r)$, producing a expected signal for the $k$th source of 
$S_k=R f(\theta_k,r) <E(r)>_k$. We assume that the noise receives Poisson contributions from both the background and the object 
counts, $N_k = \sqrt{(R f(\theta_k,r) + B \pi r^2) <E_s(r)>_k}$.  Hence the signal-to-noise is given by 

\begin{equation}
\left(\frac{S}{N}\right)_k = 
\frac{R f(\theta_k,r) \sqrt{<E(r)>_k}}
     {\sqrt{R f(\theta_k,r) + B \pi r^2}}
 \; \; \; .  
\end{equation}

For each source, we choose the value of $r$ that maximizes this function.  In our case, each individual source is dim enough that background 
fluctuations dominate the noise, and assuming that $E_{ij}$ is slowly varying, the function to be maximized is just $f(\theta_k,r)/r$. The resulting
values of $r$ as a function of off-axis angle for the soft and hard bands are shown in Figure~S\ref{opt_rad_angle}. A minimum radius of 1$''$ was assumed
in order to avoid flux losses due to astrometric problems and pixel aliasing. 

We will stack the estimated source count rates, $R_{k}$, rather than the observed source counts, $C_k$, since the latter depends on the PSF 
and exposure time at each source position but the former does not.  This is superior to performing photometry on the stacked image, which requires 
using a constant aperture for all sources.  The standard technique in the literature of performing photometry on the stacked image and 
dividing by the number of sources is both sub-optimal and biased; the bias could be fixed by instead dividing by the effective number 
of sources given the PSF and exposure information.  

While a straightforward averaging (``stack'') of the source count rates is unbiased, it is still not optimal because it gives equal weight 
to each count rate even though some were measured with lower SNR.  We can derive the weights that optimize $S/N$ of the combined stack 
by noting that the combined stack will have $S=\sum_i w_i S_i$ and $N^2=\sum_i w_i^2 N_i^2$ (since the weighted sum is just an unweighted 
sum of new objects with signal $w_i S_i$ and noise $w_i N_i$\cite{gawiser06a}).  It is easy to show that the optimal weight is then $w_i = S_i/N_i^2$ which has the 
required behavior that multiplying all the weights by a constant $k$ preserves the final $S/N$.

Then, we obtain that:

\begin{equation}
 \frac{S}{N} =  \frac{\sum_i w_i S_i}{\sqrt{\sum_i w_i^2 N_i^2}}
= \frac{\sum_i S_i^2/N_i^2}{\sqrt{\sum_i S_i^2/N_i^2}}
= \sqrt{\sum_i S_i^2/N_i^2} 
\end{equation}
Hence the optimal weights cause $S/N$ to add in quadrature so it will never 
formally decrease. However, in practice there is little benefit to including objects that 
offer individual $S/N$ of less than 10\% of the typical objects in the 
field, which is why we only stacked sources closer than 9$'$ from the average
{\it Chandra} aim point.

\end{document}